\documentclass{aa}
\usepackage{natbib}
\usepackage{amssymb}
\usepackage{amsmath}
\usepackage{graphicx}
\def\aap{A\& A}

\def\apj{ApJ}

\def\kel{\mathrm{K}}

\def\um{$\mu\mathrm{m}$}

\def\AU{\mathrm{AU}}

\def\alb{\omega}
\def\eff{\mathrm{eff}}

\def\comma{\,,}
\def\fullstop{\,.}
\def\thttle{The effect of scattering on the structure and SED of
protoplanetary disks}
\begin{document}
\title{\thttle}
\author{C.P.~Dullemond and A.~Natta}
\authorrunning{Dullemond \& Natta}
\titlerunning{Scattering in protoplanetary disks}
\institute{Max Planck Institut f\"ur Astrophysik, Karl
Postfach 1317, D--85741 Garching, Germany; e--mail:
dullemon@mpa-garching.mpg.de
\and Osservatorio Astrofisico di Arcetri,
Largo E.~Fermi 5, 50125 Firenze, Italy}
\date{DRAFT, \today}

\abstract{In this paper we investigate how the inclusion of scattering of
the stellar radiation into a passive flaring disk model affects its
structure and spectral energy distribution, and whether neglecting it could
significantly decrease the model reliability. In order to address these
questions we construct a detailed 1+1D vertical structure model in which the
scattering properties of the dust can be varied. Models are presented with
and without dust scattering, and for different albedos and phase
functions. It is found that scattering has the effect of reducing the disk
temperature at all heights, so that the disk ``shrinks", i.e., the the
density at all intermediate heights decreases. However, this effect in most
cases is more than compensated by the increase of the total extinction
(absorption + scattering) cross section, so that the surface scale height
increases, and images in scattered light will see a slightly thicker disk.  The
integrated infrared emission decreases as the albedo increases, because an
increasing part of the flux captured by the disk is reflected away instead
of absorbed and reprocessed. The reduction of the infrared thermal emission
of the disk is stronger at short wavelengths (near infrared) and practically
negligible at millimeter wavelengths. For relatively low albedo
($\alb\lesssim 0.5$), or for strongly forward-peaked scattering ($g$ roughly
$>0.8$), the infrared flux reduction is relatively small.}

\maketitle

\begin{keywords}
accretion, accretion disks -- circumstellar matter 
-- stars: formation, pre-main-sequence -- infrared: stars 
\end{keywords}

\section{Introduction}
In recent years, models of irradiated passive protoplanetary disks have
proven to be quite successful in explaining the spectral energy
distributions (SEDs) of T Tauri (TTS) and Herbig Ae/Be (HAeBe) stars (Kenyon
\& Hartmann \citeyear{kenyonhart:1987}; Calvet et
al.~\citeyear{calvetpatino:1991},\citeyear{calvetmagris:1992}; Malbet \&
Bertout \citeyear{malbetbertout:1991}; Chiang et
al.~\citeyear{chianggold:1997},\citeyear{chiangjoung:2001}; D'Alessio et
al.~\citeyear{dalessiocanto:1998}, \citeyear{dalessiocalvet:1999}; Malbet et
al.~\citeyear{malbetlachaume:2001}; Natta et
al.~\citeyear{nattaprusti:2001}; Dullemond et al.~\citeyear{duldomnat:2001};
Dominik et al.~\citeyear{domdulwatwal:2003}). Though the exact geometry of
such disks is still an issue of debate, the basic idea of a passively
reprocessing dusty disk, possibly with a remnant tenuous envelope
surrounding it, thusfar seems to withstand the test of time. However, it is
clear that analyzing the SEDs alone is in most cases not sufficient to
constrain the disk's parameters or to test theories (e.g.~Chiang et
al.~\citeyear{chiangjoung:2001}), and that spatially resolved images at
various wavelengths are required in addition.  Spatially-resolved images at
millimeter wavelengths and optical and near-infrared images in scattered
light provide valuable constraints to the disk geometry and to the
properties of the dust in these disks (see, for example, D'Alessio et
al.~\citeyear{dalessiocalvet:2001}; Testi et al.~\citeyear{testnat:2001},
\citeyear{testnat:2003} for the use of millimeter maps to constrain outer
disk properties and Cotera et al.~\citeyear{coterawhitneyyoung:2001} and
references therein for the interpretation of disk images in scattered
light).  Models that self-consistently and simultaneously compute the disk
structure, the SED and scattered-light images are required in order to fully
exploit all the observed data. Some self-consistent models of this kind
exist (e.g.~D'Alessio et al.~\citeyear{dalessiocalvet:2001}; Malbet et
al.~\citeyear{malbetlachaume:2001}), but many other models use the
assumption of zero albedo while computing the structure and SED of the disk
(e.g.~the models of Chiang et al.~and those of Dullemond et al.). Most of
the calculations of scattered light disk images use parametric disk models
to describe the disk structure (see, for example, Wood et
al.~\citeyear{woodkenwhitturn:1998}, \citeyear{woodwolfbjwh:2002}; Wolf et
al.~\citeyear{wolfpadstap:2003}). Although very successful in deriving a
first characterization of the disk surface properties, this approach does
not take into account the effects of scattering on the structure and thermal
emission of the disk.

In this paper, we explore systematically the effect on the disk structure
and emission of including scattering of the stellar radiation by the grains
in the disk. We do this by varying over a large range of values the albedo
and phase function of the grains, and we consider disks heated by a typical
Herbig Ae stars and by a typical TTS.  Our study is somewhat complementary
to D'Alessio et al.~(\citeyear{dalessiocalvet:2001}), who present disk
models where the grain size distribution is extended to very large grains
and the cross sections for absorption and scattering are changed
accordingly.  Their approach is more physical, since it relates the adopted
grain model to the specific physical process of grain growth. However, a
parametric study like ours has the advantage of greater freedom in choosing
the dust properties, and this is particularly important when one is
interested in understanding the relevance of scattering in disk models in
general.

In \S 2 
we present a self-consistent 1+1D vertical structure model for
an irradiated passive flaring disk around a TTS or a HAeBe star in
which the scattering properties of the dust grains can be varied freely. In
the case of zero albedo this model reduces to the earlier published model by
Dullemond, van Zadelhoff \& Natta (\citeyear{dulvzadnat:2002}, henceforth
DZN02). For non-zero albedo the differences to the latter model can be
investigated in detail, and an assessment can be made of the role that
scattering of the stellar radiation
plays in the structure and SED of these disks.
We illustrate the results for a disk annulus in \S 3, and for full disk models
in \S 4. A summary and conclusions follow in \S 5.

\section{The vertical structure model with scattering}
The disk vertical structure model that we present in this section is based
for most part on the 1+1D model of DZN02. The continuum radiative transfer,
which is required in order to compute the dust temperature, is done in two
stages. In stage 1, the primary stellar photons are followed as they travel
from the stellar surface outwards into the surface layers of the flaring
disk. In order to avoid having to solve a 2-D problem, this is done using a
``grazing incident angle recipe'' (also called ``flaring angle recipe'')
which splits the problem into a 1-D vertical transfer problem at each
radius. Photons are inserted into the disk at a flaring angle, which is
computed self-consistently from the disk geometry. As the photons enter the
surface layers, they get absorbed by the dust and leave their energy
behind. In the second stage, the re-emission and subsequent radiative
diffusion of this energy in the infrared regime is modeled by solving a 1-D
radiative transfer problem using the powerful variable Eddington factor
technique. This technique allows us to solve the transfer problem at any
optical depth, no matter how high. This is a major advantage over the often
used (accelerated) lambda iteration techniques. It is assumed that the disk
is passive, i.e.~there is no internal production of heat by viscosity or
any other mechanism. Therefore the above procedure suffices to determine the
dust temperature at every height above the midplane. From this, the dust+gas
density structure can be determined by integrating the equation of vertical
hydrostatic equilibrium for a given surface density $\Sigma$, under the
assumption that the gas and the dust temperatures are the same and the
gas-to-dust ratio is constant (we take it 100). The entire procedure is
iterated in order to get a self-consistent vertical temperature and density
structure. It typically requires 4 to 6 iterations for a convergence to
within 1\%. For more details on the model see DZN02.

In order to include scattering into this model, we replace stage 1 with a
Monte Carlo code capable of modeling isotropic as well as small-angle
scattering. In this way scattering of the primary stellar photons is
included in a self-consistent way. The possible scattering of the re-emitted
infrared photons can not be treated in this way, and will be ignored in the
following. In practice, our code (and the results presented here)
apply to disk atmospheres where the grains are relatively small, and
their albedo is negligible at the wavelengths of the reprocessed 
radiation (i.e., in the infrared). 

The flaring angle $\alpha$, under which the photons first enter the disk's
surface layers before they scatter, is given by:
\begin{equation}
\alpha = 0.4 \frac{R_{*}}{R} + R\frac{d}{dR}\left(\frac{H_s}{R}\right)
\comma
\end{equation}
(see e.g.~Chiang \& Goldreich \citeyear{chianggold:1997}), where $H_s(R)$ is
the surface height of the disk, defined as the vertical height $Z$ at which
the radial optical depth towards the star at stellar wavelength is
unity. This flaring angle is recomputed after each iteration, so that in the
end of the iteration procedure one obtains a self-consistent solution with
the proper flaring angle. As was described by Chiang et
al.~(\citeyear{chiangjoung:2001}), one has to employ a special method to
compute the flaring angle such that numerical instabilities do not appear.

\subsection{Monte Carlo method}\label{subsec-montcarl}
In the Monte Carlo radiative transfer we let primary stellar photons impinge
on the disk at an angle $\alpha$ with respect to the surface. We use a
vertical 1-D plane-parallel geometry, with a vertical spatial coordinate $Z$
ranging from $Z=0$ (the equator) to $Z=Z_{\mathrm{max}}$. In this approach,
the incoming photons are assumed to have $\mu_0=-\sin(\alpha)$ in the
traditional notation used in 1-D radiative transfer. For each frequency bin
we use $N$ photon packages. Each photon package is assumed to carry an
energy (per unit disk surface area) of
\begin{equation}\label{eq-input-flux-per-pack}
E^0_\nu = \frac{\alpha L_\nu^{*}} { 4 \pi R^2 N }
\fullstop
\end{equation}
In the classical Monte Carlo approach each photon package is now followed as
it scatters through the medium (randomly changing it's direction $\mu$ at
each scattering event), until it gets absorbed at some point. Absorption is
thus treated as a discrete event. By counting the number density of absorbed
photon packages per grid cell, one can determine the energy deposition into
each cell. This approach, although formally valid, has a serious weakness:
it produces enormous numerical noise at small optical depths.

A better approach is to treat absorption as a continuous process, peeling
off energy from the photon package according to the law:
\begin{equation}
\frac{dE_\nu}{ds} = - \rho \kappa^a_\nu E_\nu
\comma
\end{equation}
where $\kappa_\nu^a$ is the absorption opacity, and $s$ is the path length
along the photon trajectory which obeys the relation $ds=dz/\mu$. The energy
lost by the package is deposited in the cell in which it was lost, while the
package continues to move on until its energy $E$ has dropped below some
prescribed fraction of the original energy $E^0_\nu$, or until the package
has left the disk. Every time a photon package passes through a cell,
another portion of energy is deposited into that cell, until all photon
packages have been used up. The total energy (per unit surface area of the
disk) deposited in the cell constitutes a heating rate $q$ per unit volume.

Scattering is still treated in a discrete way, and the entire package
regularly changes direction due to these discrete scattering events.  At the
start of the random walk, and after each scattering event, a random number
$\xi_1$ uniformly distributed in the domain $\langle 0,1\rangle$ is chosen using
the random generator {\tt ran2} of Numerical Recipes
(\citeyear{numrecip:1992}). Using this number, the location of the next
scattering event is computed according to:
\begin{equation}
z_{\mathrm{new}} = z_{\mathrm{current}} - 
\frac{\mu}{\rho\kappa_s}\log\xi_1
\fullstop
\end{equation}
At this location, the Henyey-Greenstein scattering function is invoked
to compute the change in direction. This is first done in the photon's
comoving frame. The Henyey-Greenstein (1941) function is:
\begin{equation}
P(\cos\theta')=\frac{1}{2}\frac{1-g^2}{(1+g^2-2g\cos\theta')^{3/2}}
\comma
\end{equation}
where $g$ is the forward-peakedness parameter which is a property of the
dust grains, and $\theta'$ is the angle between the new photon direction and
the old one. Using again the random number generator one can determine the
$\theta'$:
\begin{equation}
\cos(\theta') = \frac{1}{2g}\left(1+g^2-
\frac{(1-g^2)^2}{(1-g+2g\xi_2)^2}\right)
\comma
\end{equation}
where $\xi_2$ is again the random number between $0$ and $1$. The other
angle ($\phi'$) is simply a random number between $0$ and $2\pi$:
$\phi'=2\pi \xi_3$. We now employ a rotation matrix operation to
rotate these angles to the real ones $\theta$, $\phi$, 
the new $\mu$ being $\mu=\cos\theta$.

Using the above procedure, one can compute the heating $q(z)$ due to the
incidence of primary stellar photons.  This heating rate then serves as the
source term for stage 2: the radiative transfer of thermal emission from the
disk.  The continuous treatment of absorption in the Monte Carlo simulation
assures that this function is virtually free of numerical noise even for a
modest number of photon packages (say, $N=100$). This method (which is
similar to the method of Van Zadelhoff et al.~\citeyear{zadelaikhodi:2003})
works very well at all optical depths. We have tested that the method
reproduces the results of a Monte Carlo code using discrete absorption
events.

\subsection{Comparing to diffuse reflection theory of Chandrasekhar}
\label{sec-chandra}
The above numerical procedure can be tested against the analytic
solutions of diffuse scattering (Chandrasekhar \citeyear{chandra:1950}).
Using so called H-functions, the problem of photons scattering off a
semi-infinite slab with a certain albedo $\alb\equiv 
\kappa_{\mathrm{scat}}/ (\kappa_{\mathrm{abs}}+
\kappa_{\mathrm{scat}})$ can be reduced to a single
numerical integral. 
For the case of isotropic scattering the $H$-function 
of Chandrasekhar is defined in the following way:
\begin{equation}\label{eq-hfunc-def}
\frac{1}{H(\mu)} = \sqrt{1-\alb} + 
\frac{\alb{}}{2}\int_{0}^{1}\frac{\mu'}{\mu+\mu'}H(\mu')d\mu'
\fullstop
\end{equation}
The values of the $H$-function can be quickly found by iteratively
evaluating Eq.(\ref{eq-hfunc-def}) until concergence is reached.  Tabulated
values can be found in Chandrasekhar (\citeyear{chandra:1950}).  Once this
$H$-function is found, one can find the reflected fraction of incident flux
$\eta$ as:
\begin{equation}
\eta = \frac{1}{2}\alb{} \int_{0}^{1}\frac{\mu'}
{\mu_0+\mu'} H(\mu_0)H(\mu') d\mu'
\comma
\end{equation}
where $\mu_0$ is the incident angle of incoming flux, assuming that the flux
is a parallel beam. In the case of a disk irradiated by a star, this
assumption is reasonably well justified. The reflection fraction $\eta$ is a
number between 0 and 1, where $\eta=1$ means that every photons is
eventually scattered away without having been absorbed, and $\eta=0$ means
that all photons have been absorbed.

\begin{figure}
\centerline{
\includegraphics[width=9cm]{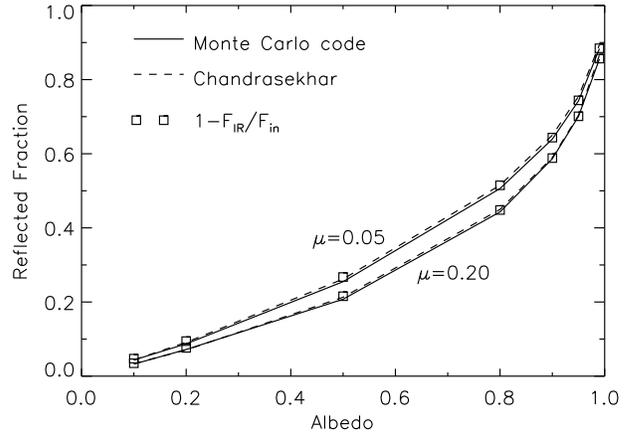}}
\caption{\label{fig-test-chandra}
The fraction of light diffusely reflected off a semi-infinite slab
as a function of albedo of the slab. Compared are the analytic results
of Chandrasekhar, the results from our Monte Carlo code, and the results
from the thermal reprocessing computed by our code.}
\end{figure}
In Fig.~\ref{fig-test-chandra} the reflected fraction of incident flux is
shown as a function of albedo. Plotted over each other are the predictions
from Chandrasekhar's theory of H-functions, the reflected fraction computed
using our Monte-Carlo code, and one minus the fraction of the flux that is
reprocessed into the infrared according to our code. All fall reasonably
well over each other, so we trust our numerical results.

Fig.~\ref{fig-test-chandra} shows that for small grazing angle and small
albedo one has $\eta\simeq \alb{}/2$, meaning that the infrared luminosity
of the disk is easily estimated to be $L_{\mathrm{IR}}\simeq
(1-\alb{}/2)L_{*}$. This result can be understood in terms of the Chiang \&
Goldreich picture of a hot surface layer: this layer intercepts all stellar
radiation, and redirects half of it away from the disk and half towards the
disk. This redirecting proceeds partly through scattering ($\alb{}$), partly
through absorption ($1-\alb{}$). This means that  a fraction of
$\alb{}/2$ of the incident flux is scattered away from the disk surface. 
The other $\alb{}/2$,
which is directed towards the disk interior, for low albedo will be almost
entirely  absorbed, 
so that $(1-\alb{}/2)$ will be reprocessed in the infrared. 
For larger albedo, the downward directed scattered photons might be
scattered once more, and escape after all. This causes the deviation from
the $\eta\simeq \alb{}/2$ estimate, as seen in Fig.~\ref{fig-test-chandra}.

\subsection{Approximate treatment of isotropic scattering}
For isotropic scattering phase function there is an interesting approximate
alternative to the full Monte Carlo approach described above (Calvet et
al.~\citeyear{calvetpatino:1991}; Strittmatter et
al.\citeyear{stritt:1974}). Using the moment equations of radiative transfer
in the Eddington approximation one can derive an analytic formula for the
mean intensity of the direct + scattered (but not the re-emitted) photon
field $J_\nu$ as a function of vertical absorption + scattering optical
depth into the disk $\tau_\nu$. If we define $\beta_\nu =
\sqrt{3(1-\alb{}_\nu)}$, and $\mu_0 =\sin(\alpha)$, where $\alpha$ is the
grazing angle of incidence of the impinging stellar photons, then the mean
intensity $J_\nu(\tau_\nu)$ can be written as:
\begin{equation}
\begin{split}
J_\nu(\tau_\nu) = & \frac{\mu_0\alb{}(2+3\mu_0)F^{*}_\nu}
{4\pi(1+2\beta_\nu/3)(1-\beta_\nu^2\mu_0^2)}e^{-\beta_\nu\tau_\nu}\\
&-\frac{3\mu_0^2\alb{}F^{*}_\nu}{4\pi(1-\beta_\nu^2\mu_0^2)}
e^{-\tau_\nu/\mu_0}
+ \frac{F^{*}_\nu}{4\pi}e^{-\tau_\nu/\mu_0}
\end{split}\label{eq-calvet}
\end{equation}
(Calvet et al.~\citeyear{calvetpatino:1991}) where $F^{*}_\nu$ is the
unextincted stellar flux at the radius of the disk annulus. Using this
expression for the mean intensity of scattered + direct stellar radiation
the heating rate $q(z)$ can be determined:
\begin{equation}\label{eq-q-calvet}
q(z) = 4\pi\int_0^{\infty} J_\nu(z) \rho \kappa_\nu d\nu
\end{equation}
This $q(z)$ then serves as the source term to stage 2 of the radiative
transfer calculation. 

We have compared the models resulting from these equations to
those resulting from the more precise Monte Carlo method described in
Sec.\ref{subsec-montcarl}. We found that under most circumstances, for
isotropic scattering phase function, the temperature structure derived from
the use of Eqs.(\ref{eq-calvet},\ref{eq-q-calvet}) and that from the Monte
Carlo method differ by less than 3\%. For isotropic scattering,
Eq.~(\ref{eq-calvet}) therefore seems to be a reliable formula. However, it
works only when the disk has large optical depth at all wavelength
where the stellar flux is important. Also, it is important to note that the
equation for the dust temperature proposed in Calvet et al.~(their Eq.14) is
not very accurate, as it is based on the use of the Rosseland mean
opacity. As was discussed by DZN02, such an approach may lead to a wrong
temperature structure.

\section{Single annulus setup}
\label{sec-eff-scat}

In order to get an understanding of the effect of scattering on the disk
structure we start with a single-annulus setup. The central star is assumed
to have the following parameters: $M_{*}=2M_{\odot}$, $R_{*}=2R_{\odot}$,
$T_{*}=10000$K (i.e.~$L_{*}=36L_{\odot}$). These are parameters relevant to
Herbig Ae stars. For simplicity we assume the spectrum of the star to be a
blackbody spectrum.  The annulus of the flaring disk we wish to study is
located at 1 AU, and has a width of 0.01 AU (i.e.~it lies between 1.00 and
1.01 AU from the central star). The grazing angle (i.e.~the angle at which
the stellar radiation enters the disk's atmosphere) is generally computed
self-consistently from the disk model, but for the simple test case
described here we assume it to be fixed to $\alpha=0.05$. The opacity is
taken to be that of a silicate grain of 0.1 $\mu$m (Draine \& Lee
\citeyear{drainelee:1984}). The scattering opacity is constructed
artificially, in order to be able to investigate different values of the
albedo. We assume that the scattering is zero for $\lambda\ge 2\mu$m, and
that it is a fixed fraction of the absorption for $\lambda<2\mu$m (keeping
the absorption opacity itself unchanged).  In this way we can freely vary
the albedo for the incident stellar radiation, while the albedo for the
reprocessed radiation remains zero. It should be noted that with this
definition the total (scattering+absorption) opacity for $\lambda<2\mu$m
increases for increasing albedo. The surface density of our test setup is
defined using the optical depth for the case of zero albedo.  The annulus
has a surface density of $\Sigma=0.3$, and (for $\alb=0$) a
visual optical depth in the vertical direction of $\tau_V=10$.

For the setup described above we compute a number of models varying the
albedo and the phase angle.  In a first set of models we assume that the
scattering is isotropic and change the albedo in the range 0.5--0.9.  In a
second set of models, we will fix the albedo to be $\omega$=0.8 and consider
forward-peaked scattering by varying the phase angle from 0.0 to values in
the interval 0.5--0.9.  In the ISM, typical values of the albedo are of
about 0.5, with $g\sim$0.5 (Draine and Lee \citeyear{drainelee:1984}; Kim et
al.~\citeyear{kimmartinhen:1994}). Both parameters, however, depend strongly
on the grain properties, such as size and chemical composition
(e.g.~Mishchenko et al.~\citeyear{mishhovtrav:2000}).

\subsection{Varying the albedo}

\begin{figure}
\centerline{
\includegraphics[width=9cm]{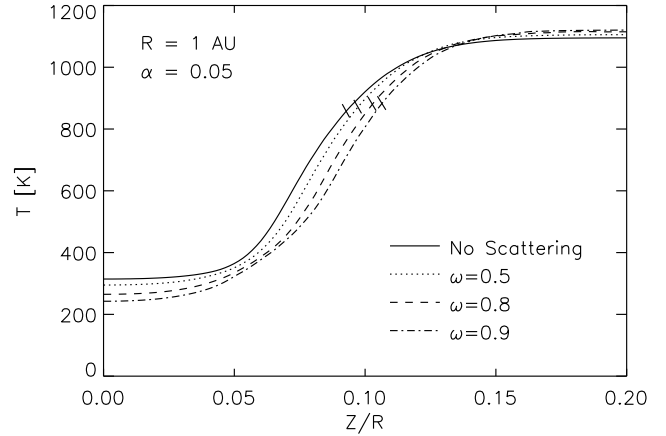}}
\caption{\label{fig-effect-verttemp} Vertical temperature profile of the
test annulus as function of $Z/R$ for different values of the albedo, as
labelled.  The tickmark on each curve indicates the position of the surface
height $H_s$ of the disk, i.e.~the height where the optical depth
(absorption + scattering) with respect to stellar photons is unity.}
\end{figure}

In Fig.~\ref{fig-effect-verttemp} the vertical temperature profile is shown
for four values of the albedo, from 0.0 to 0.9 and isotropic scattering.  As
the albedo increases, an increasing fraction of the stellar energy is lost
from the heating budget.  The effect of scattering is to decrease the disk
temperature at all heights, with the only exception of the very optically
thin region on the surface, where the stellar photons reflected away from
the disk slightly increase the mean intensity of the heating radiation.  The
temperature decrease is larger at intermediate optical depth, while the
midplane temperature drops only slightly, by 6\% for $\omega$=0.5 and a
maximum of 23\% for $\omega$=0.9. It is easy to understand why this effect
is so small if one looks at it from a two-layer perspective (Chiang \&
Goldreich \citeyear{chianggold:1997}). The disk midplane is heated by the
emission from the surface layer, part of which is thermal dust continuum
$(1-\omega)$ and part of it is scattered light ($\omega$). Only a fraction
($\omega/2$) of the downward scattered light has a chance to scatter
backwards away from the disk, and escape the fate of being absorbed and
reprocessed. This means that the effect on the midplane temperature should
be of the order of $\omega^2$ (see \S \ref{sec-chandra}).

\begin{figure}
\centerline{
\includegraphics[width=9cm]{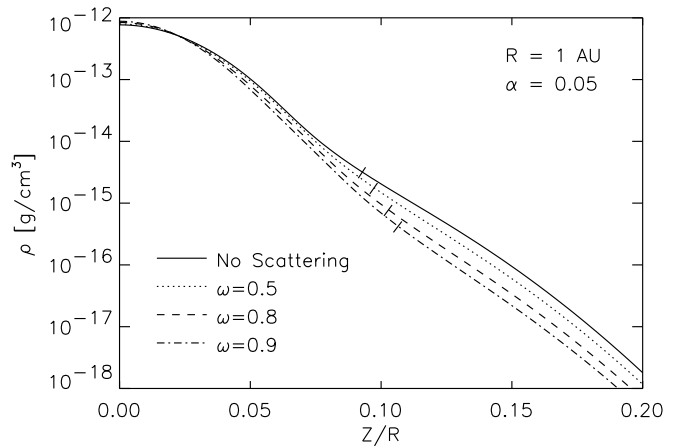}}
\caption{\label{fig-effect-vertdens}
Same as Fig.~\ref{fig-effect-verttemp} but now for the 
vertical density profile. 
}
\end{figure}

Fig.~\ref{fig-effect-vertdens} shows the corresponding density profiles. One
can see the deviation from the gaussian vertical profile expected in an
isothermal disk.  The effect is significant at high $Z$, where the increase
of the temperature due to stellar irradiation occurs. One can also see that
at fixed $Z$ the density is lower for higher values of $\omega$.  However,
the effect of scattering on the pressure scale height $H_p$ is minimal,
since $H_p$ depends on the square root of the midplane temperature, which
changes very little (see Fig.~\ref{fig-effect-verttemp}).

Fig.~\ref{fig-effect-verttemp} and Fig.~\ref{fig-effect-vertdens} show the
location of the surface height $H_s$, i.e., where the 
stellar-flux-averaged optical
depth (absorption + scattering) to the stellar photons entering the disk
under the angle $\alpha$ is unity. The location of $H_s$, which defines the
``surface'' of the disk, moves upwards as $\omega$ increases, reflecting the
increase in total optical depth as more scattering opacity is added.  Even
if for increasing albedo the density structure of the disk ``shrinks", this
effect is counteracted by the increased total opacity, and $H_s$ in fact
moves towards higher $Z/R$.

It is interesting to note that intuitively one might expect
scattering to {\em increase} the temperature of the disk just below the
surface $H_s$, since scattering redirects stellar radiation deeper into the
disk. However, as can be seen in Fig.~\ref{fig-effect-verttemp}, such an
increase in temperature does not take place. The explanation lies in energy
conservation. The deep layers of the disk get heated indirectly via emission
from the surface layer of the disk. Without scattering this downward flux is
purely thermal emission from the hot dust in the surface layer. When
scattering is included, the total flux that is beamed downward stays the
same, but now consists partly of thermal emission from the surface layer
and partly of scattered light at stellar wavelengths. The only effect that
can be discerned is that, due to the different absorption cross sections in
these two wavelength ranges, the scattered photons are absorbed higher up in
the disk than the thermal photons. This very subtle effect is seen in the
$\alb{}=0.9$ temperature curve in Fig.~\ref{fig-effect-verttemp} between
$Z/R=0.04$ and $Z/R=0.08$, where the slope changes less rapidly than for
smaller values of $\alb{}$.

\begin{figure}
\centerline{
\includegraphics[width=9cm]{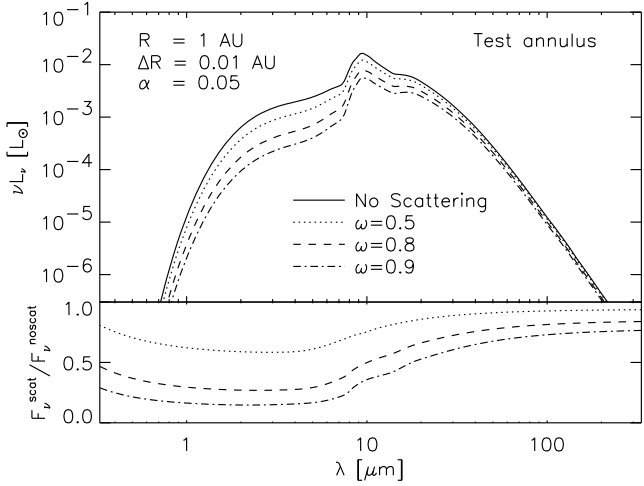}}
\caption{\label{fig-effect-specannulus} SED of the test annulus
for different values of the albedo, as labelled. The flux from the star is
not included, nor is the flux from the radiation that is scattered away
from the disk. Only the thermal emission is shown.}
\end{figure}

Fig.~\ref{fig-effect-specannulus} shows the resulting SEDs for our test
annulus for the same values of the albedo $\omega=0,0.5,0.8,0.9$. Only the
thermal emission from the annulus is shown, not the scattered-away radiation
which remains at stellar wavelengths. As one can see, the
main effect is to reduce the overall infrared flux from the annulus. This is
a logical consequence of energy conservation: as more radiation is scattered
away instead of being absorbed-and-re-emitted, the total budget of
reprocessable radiation is of course lower. In this particular example, the
fraction of reprocessed (i.e.~not scattered) stellar radiation varies from 1
for $\omega=0$ to 0.75 for $\omega=0.5$ and to 0.37 for $\omega=0.9$.

At short wavelengths the reduction is stronger than at long wavelengths, so
that the SED appears redder. At mm wavelengths the reduction between
$\omega=0$ and $\omega=0.5$ is about 6\% (proportional to the decrease of
the midplane temperature), while at $\lambda=3\mu$m the reduction is about
40\%. If one quantifies the reddening by the ratio of the NIR/IR emitted
luminosity, with NIR between 1 and 7 $\mu$m, it reduces from 0.19 for
$\omega=0$, to 0.16 for $\omega=0.5$ and 0.09 for $\omega=0.9$. The reason
for this is that the NIR emission forms in the upper layers of the disk
where scattering has a stronger effect than near the midplane. At very long
long wavelengths the disk emission is proportional to the temperature of the
midplane, which, as we have discussed, decreases only little.

\subsection{Small angle scattering versus isotropic scattering}
\label{sec-small-angle-scat}
\begin{figure}
\centerline{
\includegraphics[width=9cm]{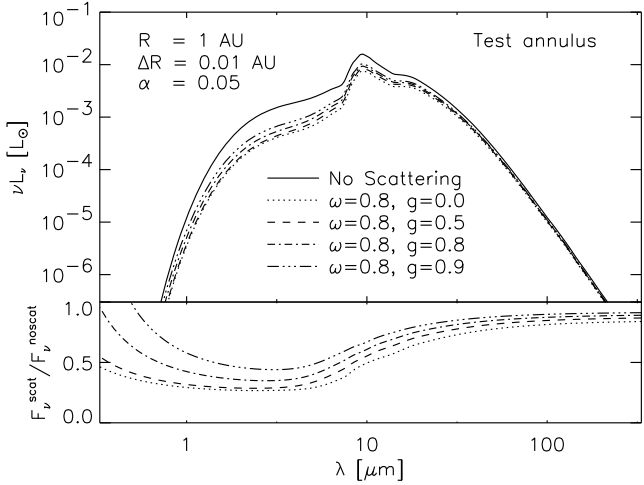}}
\caption{\label{fig-spec-effectg}
The effect of increasing the small-angle scattering factor $g$ for constant
albedo $\alb$ on the spectrum of the annulus test case.}
\end{figure}

In order to find out which effect a non-isotropic scattering phase function
has on the structure of the disk and on the SED, we take the single-slice
setup of the previous section, and make runs for different values of $g$,
using the Henyey-Greenstein phase function. For $\alpha=0.05$ and $\alb=0.8$
the differences in the SED are shown in
Fig.~\ref{fig-spec-effectg}.
From this figure it can be seen that increasing $g$ has the effect of
decreasing the influence of scattering on the disk. As the $g$ factor
increases, the infrared spectrum of the annulus also increases, since the
scattering becomes less effective in scattering radiation away before it can
thermalize and be reprocessed into the infrared. For increasing $g$, the SED
progressively moves towards the SED as computed from the case without
scattering.

The reason why this is the case can be understood by taking the extreme
example of perfectly forwardly peaked scattering ($g=1$), which is
equivalent to no scattering at all, since a scattered photon will keep
moving in the same direction as before the scattering event. For $g=1$, one
can therefore be sure that the results should be identical to the case of no
scattering, even for high albedo. For smaller $g$ scattering
becomes important, but the forward peakedness of scattering will supress the
effect of scattering compared to the case of $g=0$. This is because the
forward peakedness of scattering phase function improves the chance of still
pointing into the disk after the first scattering event, but still leaves a
chance to get scattered away.

The question now arises: could one simulate the effects of small angle
scattering by isotropic scattering with a reduced ``effective" albedo?  To
answer this we set up the following procedure. We start with a calculation
of the kind described above with a particular $\alb{}$ and $g$. Then we make
a new calculation with $g=0$, and tune the $\alb{}$ until we obtain the same
reflection fraction as in the original computation. This `effective' albedo
$\alb{}_{\mathrm{eff}}$ can be automatically found by placing the model
computation as a function-call inside a root finding routine (for instance
{\tt zbrent} from Numerical Recipes \citeyear{numrecip:1992}).  We found
that for any value of the parameters $R$, $\alpha$, $\omega$, $g$ we
explored, it is possible to find a value of $\omega$ (called
$\omega_{\mathrm{eff}}$) that reproduces the same SED with $g=0$ to
within 5\%. This effective albedo is always $\omega_{\mathrm{eff}}\le
\omega$.

\begin{figure}
\centerline{
\includegraphics[width=9cm]{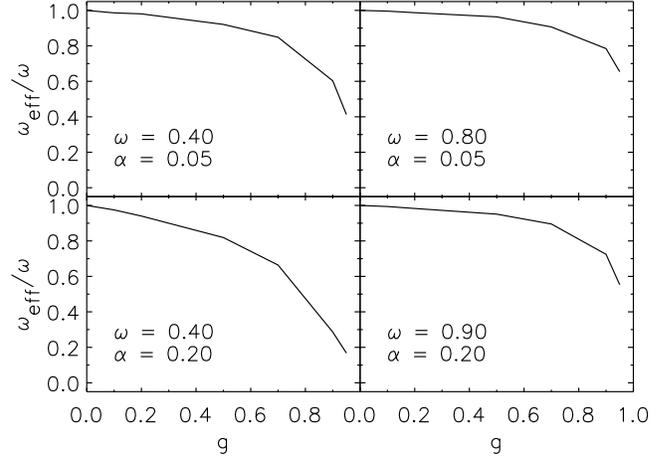}}
\caption{\label{fig-reduxion-gfact}
The effective reduction of the albedo as a result of non-isotropic 
scattering, for different values of  $\alb$ and
incident grazing angle $\alpha$.}
\end{figure}
In Fig.~\ref{fig-reduxion-gfact} the effective  reduction of the albedo
(i.e.~$\alb_{\eff}/\alb$) is shown for increasing value of $g$ for different
values of the incident grazing angle $\alpha$ and original albedo $\alb$. 
One can see that for small $\alb$ the reduction is stronger than for big
$\alb$. This can be explained by the fact that high albedo gives a photon
more than one chance to scatter and change its direction. Multiple
scattering will erase the memory of the original direction where the photon
came from. One also sees from the figure that a higher incident angle 
$\alpha$ will aggravate the reduction effect. This is because for higher
$\alpha$ the photon will need a larger scattering angle in order to be
able to escape the disk. In the extreme case of $\alpha=\pi/2$ (vertical
incident radiation) a photon will need to scatter essentially backwards
in order not to get absorbed within the disk. For this case the
counter effect of $g$ will be the strongest. For the limiting case of
$\alpha=0$, small angle scattering will essentially change nothing, since
even a small scattering angle has 50\% chance of deflecting the photon
away from the disk. 

An analytic fitting formula for the effective albedo is:
\begin{equation}
\alb_{\mathrm{eff}} = \alb{} (1-g^a)^b
\end{equation}
with
\begin{eqnarray}
a &=& 3.7\, \alb{}^{2.9} - 1.6\, \alpha^{0.42} + 2.5 \\
b &=& - 0.31\, \alb{}^{2.9} + 1.3\, \alpha^{0.42}
\end{eqnarray}
This formula has a mean deviation of 10\% with the computed values
within the domain of $0\le g \le 1$, $0.01\le \alpha\le 1$ and
$0.1\le \alb{} \le 0.9$.

\section{Full disk models}
We turn now to the determination of the effects on the SED of a full disk
ranging from the inner dust-evaporation radius out to a few hundred AU.  We
have set up a disk model for the same Herbig Ae star parameters as in
Sec.~\ref{sec-eff-scat}: mass $M_{*}=2M_{\odot}$, effective temperature
$T_{*}=10000\kel$ and luminosity $L_{*}=36L_{\odot}$.  The disk surface
density is taken to obey 
$\Sigma(R)=400\,(R/\AU)^{-1}$ g/cm$^2$
with an inner radius of 1.0 AU and outer radius 300 AU. The mass of this
disk is 
$M_{\mathrm{disk}}=0.08 M_{\odot}$.  
In these models the flaring
angle and pressure scale height are computed self-consistently, as in
DZN01. We vary the albedo from $\omega=0$ to $\omega=0.9$ and assume
isotropic scattering.  We ignore, for the moment, the emission from the
inner rim of the disk.

\begin{figure}
\centerline{
\includegraphics[width=9cm]{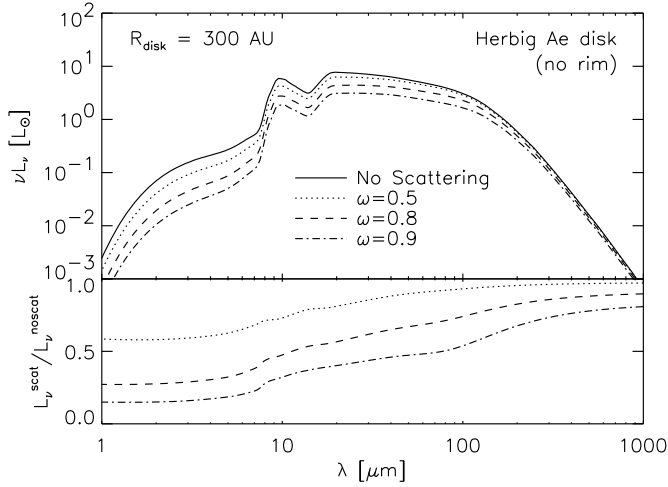}}
\caption{\label{fig-norim-albedo}
The effect of increasing albedo on the total SED of a passive flaring
irradiated disk around a Herbig Ae star.  The SED is computed at an
inclination of $i=45^{o}$. No inner rim is included. Only the
thermal emission from the disk is shown.
}
\end{figure}

The resulting SEDs are shown in Fig.~\ref{fig-norim-albedo}.  As expected
following the discussion in \S \ref{sec-eff-scat}, scattering has the effect
of reducing the infrared excess over the entire wavelength range. The ratio
of the reprocessed luminosity to the stellar one varies from 0.41 for
$\omega=0$ to 0.36 for $\omega=0.5$ to 0.19 for $\omega=0.9$. Roughly
speaking, the effect of scattering on the infrared excess is to reduce it by
a factor of the order of (1-$\omega/2$). This is not exact (see Subsection
\ref{sec-chandra}), but gives a zero order of magnitude of the effect.

As in the case of the single annulus setup the reduction is stronger at
short wavelengths than at long wavelengths. This is more clearly seen in the
lower panel of Fig.~\ref{fig-norim-albedo}, in which the reduction factor is
plotted as a function of wavelengths. The reduction is practically
negligible at millimeter wavelengths, while it is as large as 80\% (for
$\omega$=0.9) in the near infrared. It is interesting to note that the
reduction factor is a rather smooth function of wavelength, and that it
shows only a weak variation around the location of the 10 $\mu$m
silicate feature. The degree by which one sees this effect varies from
model to model. It is a subtle effect of the precise vertical profile of the
temperature and the density in the disk, and is also linked to the slight
reduction of the disk's interior temperature for increasing albedo. If we
enforce the midplane temperature of the disk to be unaffected by the
scattering (by fixing the temperature at $z=0$ to the value it had without
scattering), then the 10 $\mu$m feature turns up quite strongly in the
reduction factor.

\begin{figure}
\centerline{
\includegraphics[width=9cm]{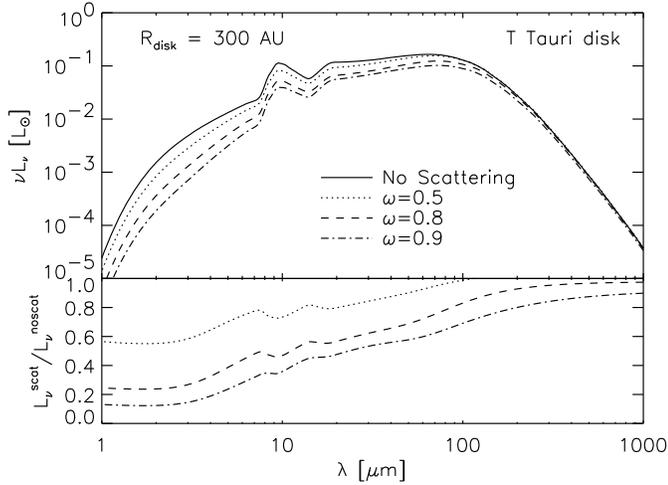}}
\caption{\label{fig-TTS-albedo}
Same as Fig.~\ref{fig-norim-albedo} for a TTS.
}
\end{figure}

\begin{figure}
\centerline{
\includegraphics[width=9cm]{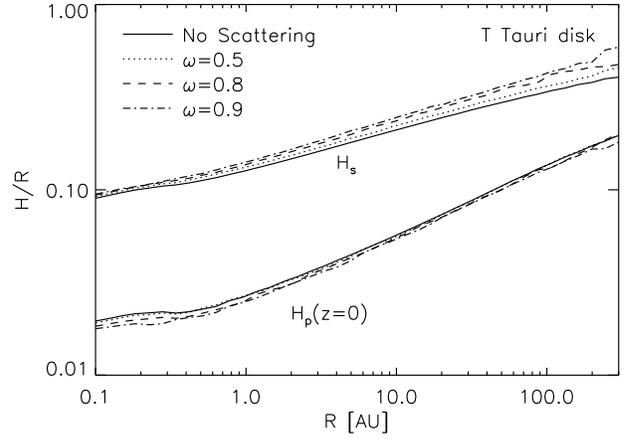}}
\caption{\label{fig-height}
Run of the pressure scale height $H_p$ and of the surface scale height
$H_s$ as function of $R$ for different values of the albedo, as
labelled. 
}
\end{figure}

Fig.~\ref{fig-TTS-albedo} shows the SEDs of disks irradiated by a T Tauri
star (TTS). The parameters of the star are taken to be: $T_{*}=4000$K,
$R_{*}=2R_{\odot}$, $M_{*}=0.5M_{\odot}$. The disk has an outer radius of
300 AU, an inner radius of $R_{\mathrm{in}}=0.1$AU, a mass of
$M_{\mathrm{disk}}=0.008\,M_{\odot}$, and a powerlaw index of the surface
density distribution of $-1$. The results are qualitatively very similar to
the Herbig Ae star case. For $\alb{}=0.5$, however, an interesting
phenomenon occurs. Longwards of 100 $\mu$m the reduction factor for the SED
exceeds unity, i.e.~an inflation instead of reduction. This can be
understood as a result of the increasing height $H_s$ of the disk, as can be
seen in Fig.~\ref{fig-height}. The grazing angle $\alpha$ increases,
enhancing the irradiation. This effect however is rather small,
and on the whole the infrared excess is reduced by the inclusion of 
scattering.

\begin{figure}
\centerline{
\includegraphics[width=9cm]{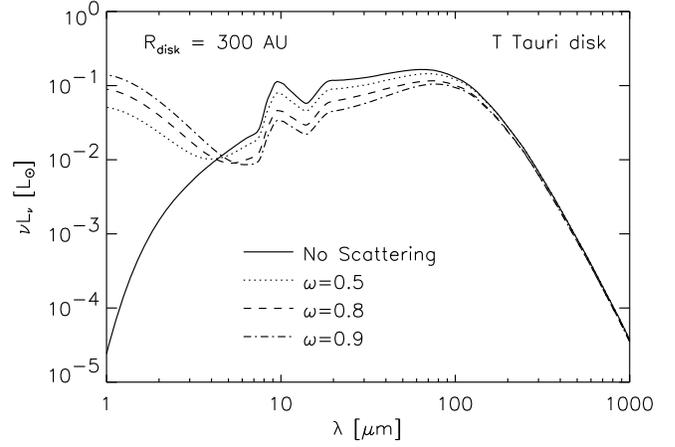}}
\caption{\label{fig-TTS-albedoscat}
Same as Fig.~\ref{fig-TTS-albedo}, but now with the scattered light 
from the disk included. The direct light from the star is not included.
}
\end{figure}

In Fig.~\ref{fig-TTS-albedoscat} the same SEDs is shown as in
Fig.~\ref{fig-TTS-albedo}, but this time the scattered light from the
disk is included in the spectrum (the direct stellar light is still
not included). As one can see: the flux of the scattered light increases
for increasing albedo. This is a combined effect of the disk becoming
more refractory and the increasing height $H_s$.

In addition to the surface height $H_s$, Fig.~\ref{fig-height} also 
shows the pressure scale height at the equatorial plane $H_p$. Contrary
to $H_s$, the pressure scale height varies only little as a function of
albedo.


\subsection{Inner rim}
For Herbig Ae/Be stars there is evidence that, in computing the SED, one
should include the effect of the stellar radiation impinging frontally on
the disk at the dust sublimation radius (Natta et
al.~\citeyear{nattaprusti:2001}; Dullemond, Dominik \& Natta
\citeyear{duldomnat:2001} henceforth DDN01). Since scattering of the stellar
light in this inner rim may be important, we have computed the SED of
rim$+$disk models for various values of the albedo.  The rim is treated in a
way similar to that of DDN01. Knowing that the rim is located at the dust
evaporation radius, we know its temperature (about 1500 K) and we can
compute the pressure scale height of the rim. The surface height of the rim
(the $z$ above the midplane at which the rim becomes transparent to
starlight) is then easily computed. As in DDN01, this height is considerably
higher than the nominal height of the flaring disk at that radius, so this
rim will in fact cast a shadow over the flaring disk.  The SED from this
disk consists then of a component in the near infrared originating from the
hot inner rim, and the emission in mid- and far-infrared originating from
the non-shadowed part of the flaring disk behind the rim. Scattering is
included in the treatment of the inner rim using the same 1-D plan-parallel
radiative transfer techniques used for the disk annuli, but this time in the
horizontal direction.

\begin{figure}
\centerline{
\includegraphics[width=9cm]{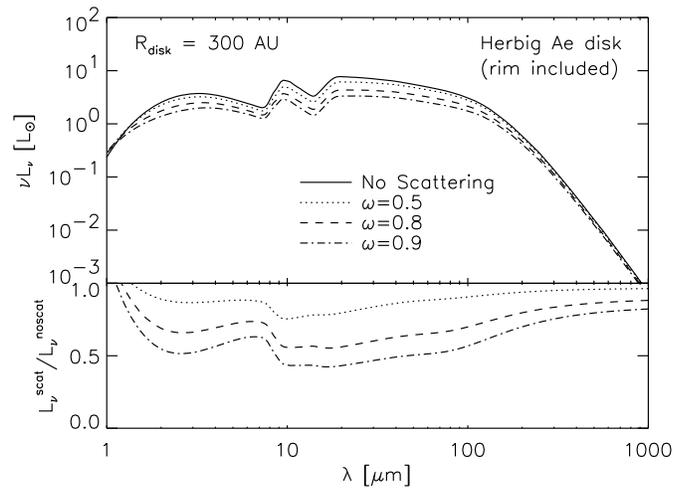}}
\caption{\label{fig-rim-albedo}
Same of Fig.~\ref{fig-norim-albedo}, but with the inner rim included.
}
\end{figure}

\begin{figure}
\centerline{
\includegraphics[width=9cm]{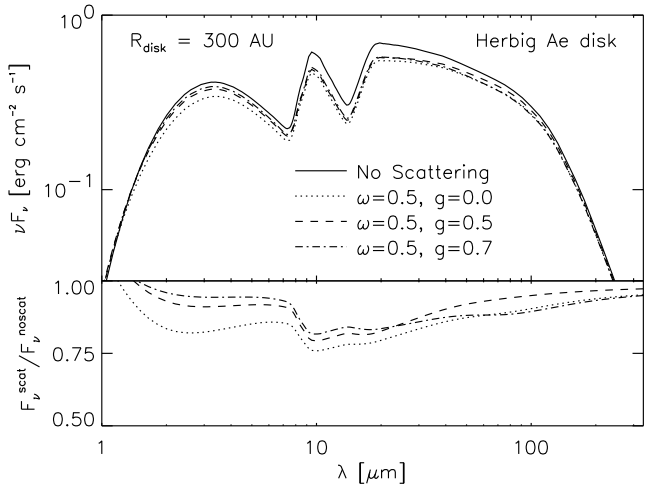}}
\caption{\label{fig-fullspec-gscat-rim} The effect of varying $g$ on the SED
of a passive flaring irradiated disk around a Herbig Ae star. In all cases
it is $\omega=0.5$. These models include the inner rim. Note that the
vertical scales are different from previous figures, in order to show the
effects more clearly.}
\end{figure}

The results are shown in Fig.~\ref{fig-rim-albedo}. The SED is very similar
to that of models with no rim at long wavelengths (in this example
approximately for $\lambda \ge 15$ \um), where the emission comes from
regions of the disk which are out of the rim shade. The region shorter of
about 7 \um\ is deeply affected by the presence of the rim, which
reprocesses in this region the intercepted stellar radiation.  The effect on
the inner rim emission of increasing $\omega$ can be seen clearly in
Fig.~\ref{fig-rim-albedo}, lower panel, and it is qualitatively similar to
what we have discussed in \S 3: the emission decreases at all wavelengths,
with the exception of the very short ones, where it is fixed by the
requirement that T=1500 K.  Globally, however, the overall shape of the SED
depends on $\omega$ less than that of models with no rim and, in particular,
the near-IR excess ($\lambda<7\mu$m) is always roughly 25\% of the total.
The total infrared excess of these models is always higher than in models
with no rim.  However, it decreases in a similar way with $\omega$, roughly
as $(1-\omega/2)$ (from 0.61 to 0.50 for $\omega=0.5$ and to 0.32 for
$\omega=0.9$).

We also made a set of rim$+$disk models for fixed albedo ($\omega$=0.5), but
varying $g$ between isotropic scattering and strongly forward-peaked
scattering (see Fig.~\ref{fig-fullspec-gscat-rim}).  Increasing the value of
$g$ decreases the effects of scattering on the SED, but it does so more
effectively for the inner rim (near-IR) than for the flaring disk
(mid-far-IR). The reason is that the inner rim is irradiated under an
incident angle of $i=90^{o}$, while the flaring disk is irradiated under a
small incident angle.  For small incident angles a photon is still likely to
skim off the surface even for large $g$, which is not the case for large
incident angle (see Sec.\ref{sec-small-angle-scat}).

\section{Summary and conclusion}
In this paper we investigate the effects of scattering of the impinging
stellar radiation on the structure and spectral energy distribution of a
passive flaring irradiated circumstellar disk. We base our model
calculations on the detailed vertical structure model of DZN02, in which
full-fledged radiative transfer is used on a density structure in vertical
hydrostatic equilibrium. It was shown in that paper that the inclusion of
full frequency-and-angle dependent radiative transfer (although in a 1+1-D
manner) can have profound effects on the vertical structure, most notably on
the dust temperature at the equatorial plane. Here we improve these models
by replacing the original treatment of irradiation by a Monte Carlo approach
that can handle small-angle scattering by dust particles\footnote{All the
models presented in this paper can be downloaded in numerical form from {\tt
http://www.mpa-garching.mpg.de/ PUBLICATIONS/DATA/radtrans/diskscat/}.}

We compared models without scattering to models with increasing albedo at
the wavelengths of the stellar radiation.  Note that we keep the absorption
cross section fixed, so that increasing albedo results in increasing the
total optical depth (scattering +absorption) for the stellar radiation.  As
the albedo increases, the disk gets cooler at all vertical heights, with the
only exception being the very upper layers. This reduction of the
temperature takes place even at the very optically thick midplane, although
the effect is smaller than at intermediate optical depth.  As a result, the
disk becomes slightly flatter (i.e.~it ``shrinks'' in vertical direction).
Still, the surface scale height $H_s$, which defines the surface where the
stellar light is intercepted, increases, since the increase of the total
opacity more than compensates for the physical disk shrinking. Disks seen in
scattered light will appear thicker than predicted by models where
scattering is ignored.

As far as the SED is concerned, increasing the albedo has the effect of
reducing the thermal emission from the disk over the entire infrared
wavelength domain. This is because scattering reflects part of the stellar
light away from the disk before it had a change to get absorbed and
reprocessed into the infrared. To zero order, for $\omega <0.5$ the
reduction is approximately $(1-\omega/2)$, and somewhat larger for larger
values of the albedo. However, we find that the spectrum is more strongly
affected at short than at long wavelengths, so that the overall SED looks
``redder".  For a typical case with $\omega=0.5$ the reduction in the near
and mid-IR could be as large as 40\%, while at mm wavelength the reduction
is of few percent at most. Models computed with different values of the
phase function show that the effect of scattering decreases strongly when it
is more forward-peaked.

In summary, we conclude that the effect of scattering of the stellar
radiation by grains in the disk has to be considered carefully in disk
models, unless the albedo is low ($\omega \ll 0.5$), the scattering is
extremely forward-peaked, or the required accuracy is not very high. 
Including scattering self-consistently in the determination of the
disk's structure is particularly important when one is interested in
predicting the disk shape that one sees in scattered light, and/or the near
and mid-infrared emission.  Only at very long wavelengths the disk
properties (both integrated flux and intensity profile) are practically
unaffected by the fraction of stellar light that is scattered rather than
absorbed by the disk.

\begin{acknowledgements}
We thank the anonymous referee for many useful comments and suggestions
which have improved the paper considerably.
We wish to thank E.~Kr\"ugel for kindly lending us his 1-D plane-parallel
radiative transfer code which we used to test our own 1+1D disk structure
code. We also thank G.J.~van Zadelhoff for discussions that proved to be
useful during the development of our Monte Carlo code. CPD acknowledges
support from the European Commission under TMR grant ERBFMRX-CT98-0195
(`Accretion onto black holes, compact objects and prototars').
\end{acknowledgements}


\begin{thebibliography}{27}
\expandafter\ifx\csname natexlab\endcsname\relax\def\natexlab#1{#1}\fi

\bibitem[{{Calvet} {et~al.}(1992){Calvet}, {Magris}, {Patino}, \&
  {D'Alessio}}]{calvetmagris:1992}
{Calvet}, N., {Magris}, G.~C., {Patino}, A., \& {D'Alessio}, P. 1992, Revista
  Mexicana de Astronomia y Astrofisica, 24, 27+

\bibitem[{{Calvet} {et~al.}(1991){Calvet}, {Patino}, {Magris}, \&
  {D'Alessio}}]{calvetpatino:1991}
{Calvet}, N., {Patino}, A., {Magris}, G.~C., \& {D'Alessio}, P. 1991, \apj,
  380, 617

\bibitem[{Chandrasekhar(1950/1960)}]{chandra:1950}
Chandrasekhar, S. 1950/1960, Radiative Transfer (New York: Dover)

\bibitem[{{Chiang} \& {Goldreich}(1997)}]{chianggold:1997}
{Chiang}, E.~I. \& {Goldreich}, P. 1997, \apj, 490, 368+

\bibitem[{{Chiang} {et~al.}(2001){Chiang}, {Joung}, {Creech-Eakman}, {Qi},
  {Kessler}, {Blake}, \& {van Dishoeck}}]{chiangjoung:2001}
{Chiang}, E.~I., {Joung}, M.~K., {Creech-Eakman}, M.~J., {Qi}, C., {Kessler},
  J.~E., {Blake}, G.~A., \& {van Dishoeck}, E.~F. 2001, \apj, 547, 1077

\bibitem[{{Cotera} {et~al.}(2001){Cotera}, {Whitney}, {Young}, {Wolff}, {Wood},
  {Povich}, {Schneider}, {Rieke}, \& {Thompson}}]{coterawhitneyyoung:2001}
{Cotera}, A.~S., {Whitney}, B.~A., {Young}, E., {Wolff}, M.~J., {Wood}, K.,
  {Povich}, M., {Schneider}, G., {Rieke}, M., \& {Thompson}, R. 2001, \apj,
  556, 958

\bibitem[{{D'Alessio} {et~al.}(2001){D'Alessio}, {Calvet}, \&
  {Hartmann}}]{dalessiocalvet:2001}
{D'Alessio}, P., {Calvet}, N., \& {Hartmann}, L. 2001, \apj, 553, 321

\bibitem[{{D'Alessio} {et~al.}(1999){D'Alessio}, {Calvet}, {Hartmann},
  {Lizano}, \& {Cant{\'o}}}]{dalessiocalvet:1999}
{D'Alessio}, P., {Calvet}, N., {Hartmann}, L., {Lizano}, S., \& {Cant{\'o}}, J.
  1999, \apj, 527, 893

\bibitem[{{D'Alessio} {et~al.}(1998){D'Alessio}, {Canto}, {Calvet}, \&
  {Lizano}}]{dalessiocanto:1998}
{D'Alessio}, P., {Canto}, J., {Calvet}, N., \& {Lizano}, S. 1998, \apj, 500,
  411+

\bibitem[{{Dominik} {et~al.}(2003){Dominik}, {Dullemond}, {Waters}, \&
  {Walch}}]{domdulwatwal:2003}
{Dominik}, C., {Dullemond}, C.~P., {Waters}, L.~B.~F.~M., \& {Walch}, S. 2003,
  \aap, 398, 607

\bibitem[{{Draine} \& {Lee}(1984)}]{drainelee:1984}
{Draine}, B.~T. \& {Lee}, H.~M. 1984, \apj, 285, 89

\bibitem[{{Dullemond} {et~al.}(2001){Dullemond}, {Dominik}, \&
  {Natta}}]{duldomnat:2001}
{Dullemond}, C.~P., {Dominik}, C., \& {Natta}, A. 2001, \apj, 560, 957

\bibitem[{{Dullemond} {et~al.}(2002){Dullemond}, {van Zadelhoff}, \&
  {Natta}}]{dulvzadnat:2002}
{Dullemond}, C.~P., {van Zadelhoff}, G.~J., \& {Natta}, A. 2002, \aap, 389, 464

\bibitem[{{Kenyon} \& {Hartmann}(1987)}]{kenyonhart:1987}
{Kenyon}, S.~J. \& {Hartmann}, L. 1987, \apj, 323, 714

\bibitem[{{Kim} {et~al.}(1994){Kim}, {Martin}, \& {Hendry}}]{kimmartinhen:1994}
{Kim}, S., {Martin}, P.~G., \& {Hendry}, P.~D. 1994, \apj, 422, 164

\bibitem[{{Malbet} \& {Bertout}(1991)}]{malbetbertout:1991}
{Malbet}, F. \& {Bertout}, C. 1991, \apj, 383, 814

\bibitem[{{Malbet} {et~al.}(2001){Malbet}, {Lachaume}, \&
  {Monin}}]{malbetlachaume:2001}
{Malbet}, F., {Lachaume}, R., \& {Monin}, J.-L. 2001, \aap, 379, 515

\bibitem[{Mishchenko {et~al.}(2000)Mishchenko, Hovernier, \&
  Travis}]{mishhovtrav:2000}
Mishchenko, M., Hovernier, J., \& Travis, L., eds. 2000, Light scattering by
  non-spherical particles (Academic Press)

\bibitem[{{Natta} {et~al.}(2001){Natta}, {Prusti}, {Neri}, {Wooden}, \&
  {Grinin}}]{nattaprusti:2001}
{Natta}, A., {Prusti}, T., {Neri}, R., {Wooden}, D., \& {Grinin}, V.~P. 2001,
  \aap, 371, 186

\bibitem[{Press {et~al.}(1992)Press, Teukolski, Vetterling, \&
  Flannery}]{numrecip:1992}
Press, W., Teukolski, S., Vetterling, W., \& Flannery, B. 1992, Numerical
  Recipes in Fortran, Second Edition (Cambridge University Press)

\bibitem[{{Strittmatter}(1974)}]{stritt:1974}
{Strittmatter}, P.~A. 1974, \aap, 32, 7

\bibitem[{{Testi} {et~al.}(2001){Testi}, {Natta}, {Shepherd}, \&
  {Wilner}}]{testnat:2001}
{Testi}, L., {Natta}, A., {Shepherd}, D.~S., \& {Wilner}, D.~J. 2001, \apj,
  554, 1087

\bibitem[{{Testi} {et~al.}(2003){Testi}, {Natta}, {Shepherd}, \&
  {Wilner}}]{testnat:2003}
---. 2003, submitted

\bibitem[{{van Zadelhoff} {et~al.}(2003){van Zadelhoff}, {Hogerheijde}, \& van
  Dishoeck}]{zadelaikhodi:2003}
{van Zadelhoff}, G.-J.~Aikawa, Y., {Hogerheijde}, M.~R., \& van Dishoeck, E.
  2003, \aap, 397, 789

\bibitem[{Wolf {et~al.}(2003)Wolf, Padgett, \& Stapelfeldt}]{wolfpadstap:2003}
Wolf, S., Padgett, D., \& Stapelfeldt, K. 2003, astroph/0301335

\bibitem[{{Wood} {et~al.}(1998){Wood}, {Kenyon}, {Whitney}, \&
  {Turnbull}}]{woodkenwhitturn:1998}
{Wood}, K., {Kenyon}, S.~J., {Whitney}, B., \& {Turnbull}, M. 1998, \apj, 497,
  404

\bibitem[{{Wood} {et~al.}(2002){Wood}, {Wolff}, {Bjorkman}, \&
  {Whitney}}]{woodwolfbjwh:2002}
{Wood}, K., {Wolff}, M.~J., {Bjorkman}, J.~E., \& {Whitney}, B. 2002, \apj,
  564, 887

\end{thebibliography}
\end{document}